\documentclass{aa}
\usepackage{graphicx}
\begin{document}
   \title{Century scale persistence in longitude distribution: in the
Sun and {\it in silico}}

   \author{J. Pelt
          \inst{1,2}
          \and
          I. Tuominen
          \inst{2}
          \and
          J. Brooke
          \inst{3,4}
          }

   \offprints{J. Pelt}
   \authorrunning{J. Pelt et al.}
   \titlerunning{Century scale persistence in longitudes}	
   \institute{
             Tartu Observatory, 61602 T\~{o}ravere, Estonia
             \and
             Astronomy Division, Department of Physical Sciences,
             P.~O.~Box~3000, FIN--90014 University of Oulu, Finland
             \and
             Manchester Computing, University of Manchester, Oxford Road, 
             Manchester, M13 9PL, UK 
             \and
             Department of Mathematics, University of Manchester, Oxford Road, 
             Manchester, M13 9PL, UK             
             }

   \date{Received ---; accepted ---}

   \abstract{
   Using Greenwich sunspot data for 120 years it was recently observed that 
   activity regions on the Sun's
   surface tend to lie along smoothly changing longitude
   strips 180\degr  apart from each
   other.
   However, numerical experiments
   with random input data show that most, 
   if not all, of the
   observed longitude discrimination can be looked upon as an artifact of
   the analysis method. 
   }

   \maketitle
%

\section{Introduction}
The time distribution of sunspot latitudes is well known and the ubiquitous
butterfly diagram is becoming a kind of trademark in solar research.
The same is not true for longitudes.
It is still not known how persistent in time 
are different
statistical features of longitude distributions for
various activity indicators 
(see for instance
Bai~\cite{Bai03} and references therein).
The present letter is inspired by the 
recent contribution by 
Berdyugina \& Usoskin~\cite{Berd03} (hereafter BU). By using certain
data processing techniques they built smooth curves which run along
central longitudes  of sunspot activity centres. They observed 
that distribution
of activity centre phases (if computed against
mean flow) has a double peak distribution at a
high level of confidence (see Fig.5 in BU). They concluded
that there is a century scale persistent feature in the sunspot longitude
distribution: the local neigbouring maxima of activity tend to 
lie 180\degr apart.
They also observed several ``flip-flop'' events
(see Jetsu et al.~\cite{Jetsu93}) and concluded that 
there is a strong analogy between 
the Sun and
rapidly rotating magnetically active 
late--type stars (for an overview of active stars see 
Tuominen et al.~\cite{Tuominen2002}). 
We can also formulate their result in metaphoric
terms: the Sun has 
two faces,
and as a kind of
peculiar Janus it sometimes switches them back and forth.

In this paper we analyse {\em randomly
generated} distributions to check the possibility that the observed 
effect in BU
is a result of dependencies and 
correlations hidden in the statistical
method itself. And indeed, we found 
that it is possible to obtain
quite similar double-peak distributions for random data. Consequently
the statistically significant part (or even all)  of the 
effect described in 
BU can be safely ascribed to the artifacts of data processing.
It is still reasonable to assume that the mean magnetic field 
on the Sun (at least in the regions of spot formation) is axisymmetric,
when the mean is taken over sufficently long time.

What follows is a short technical account of the analysis done.

\subsection{Simulations for active longitude distributions}
The strongest argument in BU for the century scale persistence
of active longitudes is given in their Fig.~5 where
the two strong and symmetric
distribution maxima indicate that sunspot groups tend to concentrate around
two centres, 180\degr\ ($0.5$ in phase) apart.

To check their result and its methodological underpinnings we did some very
simple numerical simulations\footnote{Full source codes of relevant software can be
downloaded from {\it http://www.aai.ee/$\sim$pelt/soft.htm}.}.
Instead of using actual observational data we generated
artifical random data sets and tried to process the obtained distributions
along the same lines as in the original paper. There was no need
to simulate all data processing steps performed in BU. 
It was enough to start from
the point where for each Carrington rotation the phases for activity
maxima were computed. For simplicity we assume that for
every rotation there are at least two activity maxima,
each representing a sunspot group concentration.

Each individual
simulation run proceeded then as follows.
First, for each of the $N=1720$ Carrington rotations 
with starting times $t_i, i=1,\dots,N$
two {\em fully random and statistically independent} 
phases $\phi_1(t_i),\phi_2(t_i), i=1,\dots,N$ were generated. 
The evenly distributed values \hbox{$0<=\phi_1(t_i)<1$ and $0<=\phi_2(t_i)<1$}
were then used as simulated
phases for principal and secondary maxima of sunspot density 
distribution along
longitude for particular rotation $t_i$.

\begin{figure}
   \resizebox{\hsize}{3.3 cm}{\includegraphics{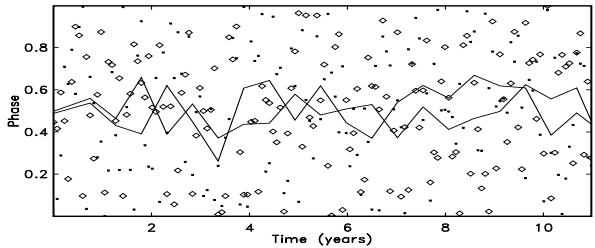}}
   \caption{Fragment of the computer generated phases $\phi_1(t_i)$ (diamonds)
   and $\phi_2(t_i)$ (points).
   Thick line is a half year average
   of the principal maxima ($M_1(t)$), thin line is a half year average
   of the secondary maxima ($M_2(t)$). The mean curves can cross each
   other.}
   \label{fg0000fig}
\end{figure}

\begin{figure}
   \resizebox{\hsize}{3.3 cm}{\includegraphics{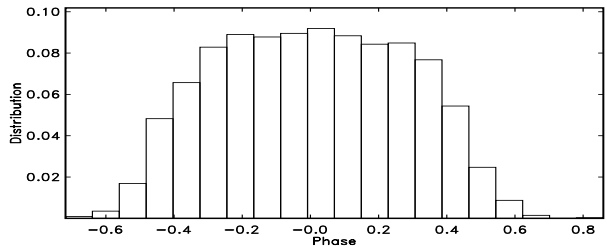}}
   \caption{Distribution of the differences between
   phases $\phi_1(t_i), \phi_2(t_i)$ and the half year mean $M^1(t)$ for
   one particular run (first from the series of 100).}
   \label{sm0000fig}
\end{figure}

For both sequences of phases $\phi_1$ and $\phi_2$ the half year means
were calculated and 
from the mean values, by linear interpolation, the two contiuous mean curves
$M_1(t)$ and $M_2(t)$ were built.
To get the analogue of the Fig.5 of BU 
the distribution of the differences
of the all generated phases
and one of the continuous curve ($M_1(t)$) was computed.
In Fig.~\ref{fg0000fig} a short fragment of generated phases
together with two running means and in Fig.\ref{sm0000fig}
the corresponding distribution for all differences
are plotted.
\begin{figure}
   \resizebox{\hsize}{3.3 cm}{\includegraphics{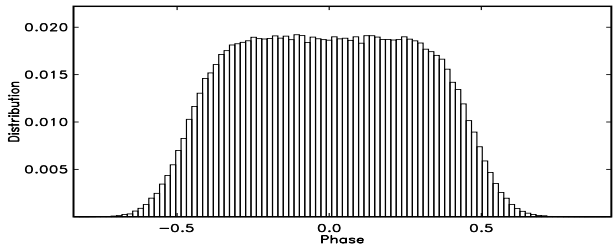}}
   \caption{Mean distribution of the $100$ separate runs.
   The distribution is flat in the centre.}
   \label{sc0000fig}
\end{figure}
To eliminate the effect of the statistical fluctuations
we generated 100 independent distributions and
in Fig.~\ref{sc0000fig} the mean distribution of these runs 
are plotted.
As we see the means are uncorrelated and the distribution
is approximately flat in its central part.

\subsection{Phase adjustments}
In BU, the original phases for active longitude
centers were not used straightforwardly, but were modified.
The authors say ``We plot the recovered phases
of sunspot clusters versus time and find that regions
migrate in phase as rigid structures. When a region
reaches $\phi=1$, it appears again near $\phi=0$.
In such cases we add an integer to the phase and
unfold continuous migration of the regions''. Unfortunately
the desription of the actual procedure of the described processing step
remains somewhat obscure. 
From Usoskin (\cite{Usoskin03}) we understood that
what was done was something similar to the following:
\begin{itemize}
\item There was a preset
constant, say $\delta \phi$.
\item When particular maximum occured at a phase which exceeded $1-\delta \phi$
and when one of the maximum phases for the next rotation was less 
than $\delta \phi$, then
such situations were looked upon as a candidate for a phase jump.
\item If the prospective phase jump improved
``continuity'' of the assumed migration paths it was accepted. 
\end{itemize}
It is important to note that the jumps were not constrained by the
type of maxima (main or secondary). To ensure 
continuity of the migration paths the authors of BU allowed swaps between
the principal and secondary maxima (``flip-flops'').

To avoid any subjective judgement we applied a similar but {\em fully automatic}
procedure
in our simulation program. Because the actual value for 
$\delta \phi$ (if it ever
existed in numerical form) was not communicated to as, we tried 
multiple values
of it. In a ``crossing situation'' we constructed all possible jumping 
schemes (no jump at all; largest phase jumps; if both 
can jump, then smaller phase jumps) and 
selected in each situation a scheme with a smallest
change in phases. 
Thus, if before crossing point, the phases
for two maxima were $\phi_1$ and $\phi_2$ and after crossing $\phi_3$
and $\phi_4$, then the selection criterion was the 
minimization of the $|\phi_1-\phi_3|+|\phi_2-\phi_4|$. We believe
that the actual procedure of the original authors 
was carried out similarily,
even if it was not implemeted as a rigorous algorithm but 
was done by manual adjusting. 

\begin{figure}
   \resizebox{\hsize}{3.3 cm}{\includegraphics{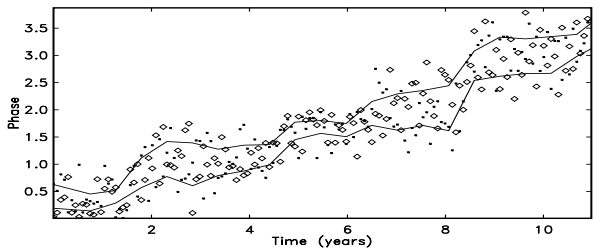}}
   \caption{Two rows of random phases with
   phase adjustments applied. The upward trend is due to the
   accumulation of the added full cycles.}
   \label{fg1100fig}
\end{figure}

We computed analogously to Fig.~\ref{sc0000fig} 
distribution diagrams for several values of $\delta \phi$
in the range $0.05-0.25$. All diagrams had a characteristic bimodal
form with two distribution maxima (persistent longitudes!). 
For presentation purposes we give here an example with
one particular parameter value $\delta \phi = 1/18$. This is the most
natural value for the case when the histogram method of activity
smoothing is used and the number of histogram bins is $18$ (see BU). 
In Fig.~\ref{fg1100fig} we see a short (approximately one solar cycle)
fragment of the two mean curves with upward drifting phases.
\begin{figure}
   \resizebox{\hsize}{3.3 cm}{\includegraphics{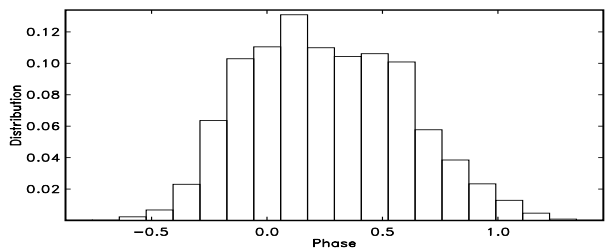}}
   \caption{Distribution of the differences between
   phases and the mean $M_1(t)$  after adjustment. 
   Crossing parameter $\delta \phi = 1/18$.
   The bimodal structure of the distribution demonstrates that century
   scale persistence of the active longitudes 
   appear even in
   randomly distributed phases.}
   \label{sm1100fig}
\end{figure}   
\begin{figure}
   \resizebox{\hsize}{3.3 cm}{\includegraphics{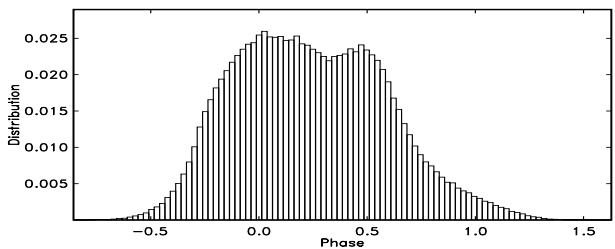}}
   \caption{Mean distribution of the differences for $100$ runs.
   Crossing parameter $\delta \phi = 1/18$.
   The bimodality in Fig.~\ref{sm1100fig} is not 
   an exceptional
   fluctuation.}
   \label{sc1100fig}
\end{figure}   
In Fig.~\ref{sm1100fig} the one concrete distribution (effectively
the first from series of 100) is given. The Fig.~\ref{sc1100fig} depicts
the mean distribution of all runs. As we see, the distributions
are of two-peaked nature. 

We started from random and uncorrelated
pairs of phases. How it is then possible that there are now certain
preferred positions for the random phases? 
It is not
hard to see that the selection of phase jumping
points tends to introduce 
extra statistical dependencies between 
neigbouring maxima in different rotations.
Some of the extra large differences $\phi(t_i)-\phi(t_{i+1})$ 
can be excluded by doing jumps for appropriate
phases. As a result
overall mean of differences between sequential phases
tends to be less. When these extra 
correlations and allowed swaps between principal and secondary minima
are combined the 
double-peaked distribution results.

\subsection{Restrictions from continuity of distributions}
For further discussion we need to introduce the notion of
circular distance $D(\phi_1,\phi_2)$. It is defined as 
$\min(|\phi_1-\phi_2|,1-|\phi_1-\phi_2|)$ and it takes into
account the circular nature of the phases (the point $1$ is the 
starting point $0$ for the next turn).

Our simulation model 
has been unrealistic up to this point. It is possible
that both randomly generated phases $\phi_1(t_i)$ and $\phi_2(t_i)$ are equal or very
near to each other. For real data this can not be so. Depending 
on the smoothing and maximum finding method, there is a certain 
minimal circular distance $\Delta \phi$ between two phases of maxima. 
To simulate this kind
of more realistic data we generated random phases 
as was done before. However for each phase pair we computed the circular 
distance $D(\phi_1(t_i),\phi_2(t_i))$ and 
pairs whose phases were too close
($D(\phi_1(t_i),\phi_2(t_i))<=\Delta \phi$) were excluded from the resulting 
data set. 
In this way we allow certain statistical
dependencies between two phases but phase pairs remain totally
independent (from rotation to rotation).

\begin{figure}
   \resizebox{\hsize}{3.3 cm}{\includegraphics{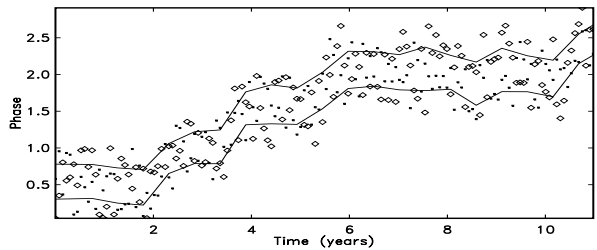}}
   \caption{Fragment of random phases for constrained data.
   The series of maxima were generated by checking against 
   minimum allowed distance between main and secondary maximum.
   Crossing parameter $\delta \phi = 1/18$ and 
   distance parameter $\Delta \phi = 0.30$.
   The adjustments to follow phase drift were also applied. 
   }
   \label{fg1110fig}
\end{figure}

\begin{figure}
   \resizebox{\hsize}{3.3 cm}{\includegraphics{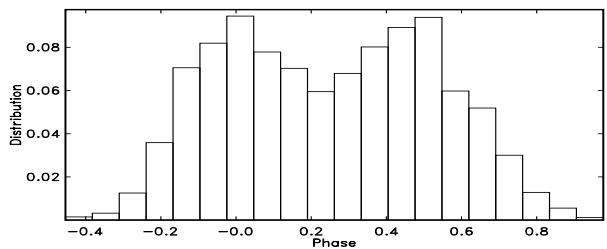}}
   \caption{
   Phase differences from mean $M_1(t)$ for
   constrained data.
   }
   \label{sm1110fig}
\end{figure}

\begin{figure}
   \resizebox{\hsize}{3.3 cm}{\includegraphics{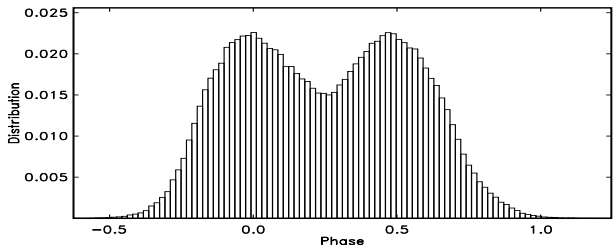}}
   \caption{Mean distribution of the phases for
   distributions like in Fig.~\ref{sm1110fig}.
   }
   \label{sc1110fig}
\end{figure}

A short fragment of data which is checked 
against minimal $\Delta \phi$ and with adjusted phases
is given in Fig.~\ref{fg1110fig} (crossing parameter 
$\delta \phi = 1/18$, distance parameter $\Delta \phi = 0.3$). 
The tendency of phases to cluster strongly
around mean curves is well demonstrated in Fig.~\ref{sm1110fig}. The
plot of the mean distribution for 100 separate runs (Fig.~\ref{sc1110fig})
shows that we are not dealing here with random fluctuations.  
It is possible to make the ``persistency'' 
effect even stronger if we select certain ``optimal''
input parameters $\delta \phi$ and $\Delta \phi$. 
\subsection{Local correlations}
If we return now to the original setup of the generation mechanism for
the random data sets, we recall that the distributions for all Carrington
rotations were assumed to be statistically independent. This is not of course
the case for real distributions. It is well known that there 
are somewhat persistent activity
regions on the Sun. 
If we try to model this kind of local correlation,
the separation of the two maxima becomes even more pronounced.

\begin{figure}
   \resizebox{\hsize}{3.3 cm}{\includegraphics{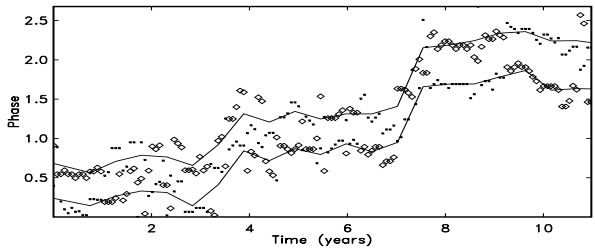}}
   \caption{Two rows of random phases and corresponding 
   mean curves for locally correlated data.
   Crossing parameter $\delta \phi = 1/18$, distance
   parameter $\Delta \phi = 0.3$.
   In this kind of simulation even ``flip-flop'' events
   can be seen.
   }
   \label{fg1111fig}
\end{figure}   
\begin{figure}
   \resizebox{\hsize}{3.3 cm}{\includegraphics{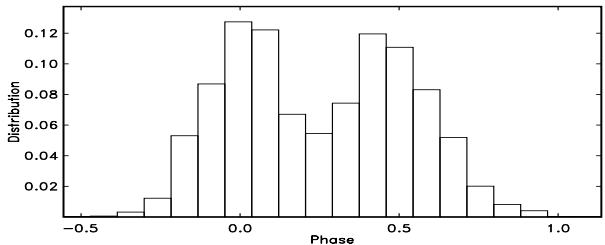}}
   \caption{
   Phase differences from mean $M_1(t)$ for
   locally correlated data.}
   \label{sm1111fig}
\end{figure}   
\begin{figure}
   \resizebox{\hsize}{3.3 cm}{\includegraphics{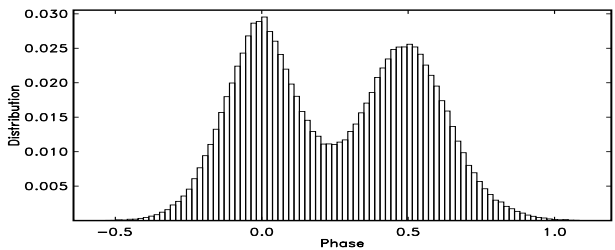}}
   \caption{Mean distribution of the phase differences for
   distributions like in Fig.~\ref{sm1111fig}.
   }
   \label{sc1111fig}
\end{figure}   

Look at the 
Figures \ref{fg1111fig} through \ref{sc1111fig}  which were obtained using locally correlated
data sets. The 
two-peaked nature 
is now very pronounced
and corresponding plots tend to be quite similar to the
plots in the original paper. By fine tuning of the parameters it is possible
to get a perfect match. But this is not our goal here.

The local correlations were introduced by using the following
generation scheme. For every new phase to be generated,
one from the four possible actions was selected (with equal probability):
\begin{itemize}
\item generate totally new phase,
\item repeat previous phase,
\item add to previous phase a small amount $\delta$,
\item subtract from previous phase a small amount $\delta$.
\end{itemize} 
As a result we got a sequence of globally uncorrelated (due to the
first option) but locally correlated random walk 
(due to the other options) fragments. We are not arguing here that
this is a very good model for a solar activity, our aim is just to
show that distributions like these in Fig.~5 of BU can
be obtained from locally correlated random sequences and that the 
correlatedness can strengthen distribution peaks of preferred longitudes.

\section{Discussion and conclusion}
The analysis above shows that phase distributions of
various parameters (maxima in our case) tend to be very
sensitive to hidden dependencies and correlations.
It is not very easy to see from the first glance that
if we look at random primary and secondary maxima, then a
phase adjusting to follow ``rigid patterns'' introduces
a significant hidden statistical dependencies which show up as double peaked
distribution of phase differences.

When we combine all three types of dependencies and 
correlations, from phase tracking,
from constrained phase differences and from 
real short time correlations, then we
easily get significantly bimodal diagrams like Fig.5 in BU. The flexibility
due to the possibility to swap principal and secondary 
maxima (``flip-flop'') and adjust
phases (``rigid patterns'') allows one to amplify every local short range
correlation into magnificent century scale persistent phenomena.

It is important to note that our simulation analysis does not rule out that
in principle there can be certain persistent phenomena in sunspot
longitude distributions. We just demonstrated that current evidence is not
sufficient.

\begin{acknowledgements}
      We are thankful to S.V. Berdyugina and I.G. Usoskin for additional
      comments about data processing procedures used in the original
      paper.
      Part of this work was supported by the Estonian Science Foundation
      grant No. 4697 and Academy of Finland grant No. 43039. 
\end{acknowledgements}

\end{document}